\documentclass[aps,twocolumn,superscriptaddress]{revtex4-1}
\usepackage[colorlinks=true,bookmarks=true,citecolor=magenta,urlcolor=magenta,linkcolor=magenta,breaklinks]{hyperref} 
\usepackage{breakurl}
\usepackage{sidecap}
\usepackage{amssymb}
\usepackage{hhline}
\usepackage{urwchancal}
\usepackage{xcolor}
\sidecaptionvpos{figure}{t}
\usepackage{amsmath}
\usepackage{graphicx}
\usepackage{esint}
\usepackage{epstopdf}
\usepackage{rotating}
\graphicspath{{pict/}{./}}
\usepackage{bm}%
\usepackage{microtype,bm,bbm,graphicx,booktabs,times}

\newcounter{Fig}

\newcommand\mymapstol{\mathrel{\ooalign{$\leftarrow$\cr%
  \kern1.75ex\raise0.275ex\hbox{\scalebox{1}[0.4]{$\mid$}}\cr}}}

\newcommand\mymapstor{\mathrel{\ooalign{$\rightarrow$\cr%
  \kern-.15ex\raise.275ex\hbox{\scalebox{1}[0.4]{$\mid$}}\cr}}}
\begin{document}


\title{Arbitrary Polarization-Independent Backscattering or Reflection by Rotationally-Symmetric Reciprocal Structures}
\author{Weijin Chen}
\email{Authors contributed equally to this work.}
\affiliation{School of Optical and Electronic Information, Huazhong University of Science and Technology, Wuhan, Hubei 430074, P. R. China}
\author{Qingdong Yang}
\email{Authors contributed equally to this work.}
\affiliation{School of Optical and Electronic Information, Huazhong University of Science and Technology, Wuhan, Hubei 430074, P. R. China}
\author{Yuntian Chen}
\email{yuntian@hust.edu.cn}
\affiliation{School of Optical and Electronic Information, Huazhong University of Science and Technology, Wuhan, Hubei 430074, P. R. China}
\affiliation{Wuhan National Laboratory for Optoelectronics, Huazhong University of Science and Technology, Wuhan, Hubei 430074, P. R. China}
\author{Wei Liu}
\email{wei.liu.pku@gmail.com}
\affiliation{College for Advanced Interdisciplinary Studies, National University of Defense
Technology, Changsha, Hunan 410073, P. R. China}

\begin{abstract}
We study the backward scatterings of plane waves by reciprocal scatterers and reveal that $n$-fold ($n\geq3$) rotation symmetry is sufficient to secure invariant backscattering for arbitrarily-polarized incident plane waves. It is further demonstrated that the same principle is also applicable for infinite periodic structures in terms of reflection, which simultaneously guarantees the transmission invariance if there are neither Ohmic losses nor extra diffraction channels. At the presence of losses, extra reflection symmetries (with reflection planes either parallel or perpendicular to the incident direction) can be incorporated to ensure simultaneously the invariance of transmission and absorption. The principles we have revealed are protected by fundamental laws of reciprocity and parity conservation, which are fully independent of the optical or geometric parameters of the photonic structures. The optical invariance obtained is intrinsically robust against perturbations that preserve reciprocity and the geometric symmetries, which could be widely employed for photonic applications that require stable backscatterings or reflections.
\end{abstract}

\maketitle

\section{Introduction}
\label{section1}

The study of backscatterings (for finite scatting bodies) or reflections (for extended infinite structures that could be homogeneous, periodic or quasi-periodic) constitutes one of the oldest and most fundamental topics in photonics that spawns many practical applications~\cite{Bohren1983_book,YARIV_2006__Photonics,JOANNOPOULOS_2008__Photonic,BARRON_2009__Molecular}.  In the past three decades, largely due to the explosive developments of photonic crystals, metamaterials, and several other sibling disciplines, this topic has attracted surging renewed interest~\cite{YARIV_2006__Photonics,JOANNOPOULOS_2008__Photonic,ZHELUDEV_metamaterials_2012}. In particular, the introduction of emerging new concepts from those fields (including optically-induced magnetism, generalized Kerker scattering, bound state in the continuum, optical topology and non-Hermiticity, and parity-time symmetry) has significantly broadened the horizons and rendered unprecedented flexibilities for efficient manipulations for backward scatterings or reflections~\cite{Pendry1999_ITMT,jahani_alldielectric_2016,KUZNETSOV_Science_optically_2016,Kerker1983_JOSA,LIU_2018_Opt.Express_Generalized,HSU_Nat.Rev.Mater._bound_2016,
OZAWA_2018_ArXiv180204173,PANCHARATNAM_1955_ProcIndianAcadSci_propagation,BERRY_CURRENTSCIENCE-BANGALORE-_pancharatnam_1994,FENG_2017_Nat.Photonics_NonHermitiana,EL-GANAINY_2018_Nat.Phys._NonHermitian,HORSLEY_Nat.Photonics_spatial_2015,IM_2018_Nat.Photonics_Universal}.
Despite those great achievements, there has been an almost pervasive limitation for investigations concerning backscattering or reflection: the studies have been largely confined to incident waves of either  circular (for
chiroptical studies in particular~\cite{BARRON_2009__Molecular}) or linear polarizations. While for practical applications, arbitrary polarization-independent responses are essential for stable operations against polarization fluctuations induced by perturbations, systematic discussions of which are vitally important while unfortunately rare.

Recently we have shown comprehensively how to exploit simultaneously rotation (or electromagnetic duality) and reflection symmetries to deliver invariant scattering properties of reciprocal optical scatters~\cite{YANG_2020_ArXiv200613466Phys._Symmetry,YANG_ArXiv200701535Phys._Scatteringa} for arbitrary polarizations covering the whole Poincar\'{e} sphere~\cite{YARIV_2006__Photonics}. Nevertheless, the invariance obtained concerns only the integrated scattering properties of cross sections (including extinction, absorption and scattering cross sections), and it has not been discussed specifically whether or not scatterings along fixed angles (backward scattering for example) are invariant or not. Moreover, it is well known that optical responses of periodic photonic structures are dictated by the unit-cell particle scatterings (lattice coupling taken into consideration) along the corresponding out-coupling directions: \textit{e.g.} forward scatterings for transmissions; backward scatterings for reflections; and scatterings along diffractive directions for diffractions of different orders)~\cite{LIU_2017_ACSPhotonics_Beam,CHEN_2019__Singularities}.  As a result, to discuss the scattering invariance along some fixed directions for arbitrary polarizations is of great significance for both finite scatterers and extended periodic structures.

In this study we discuss the backscatterings of arbitrarily-polarized incident plane waves by reciprocal scattering bodies. It is revealed that sole $n$-fold ($n\geq3$) rotation symmetry  can ensure invariant backward scattering irrespective of either optical or geometric parameters (rotation axis parallel to the incident direction). Similarly, reflection invariance can be also obtained for rotationally-symmetric periodic structures, which is accompanied by invariant transmission when there are neither Ohmic losses nor extra diffractions. At the presence of losses, it is further demonstrated that extra reflection symmetries (reflection planes either parallel or perpendicular to the incident direction) can be introduced to produce invariant absorptions and transmissions. The invariant responses obtained are generically protected by reciprocity and geometric symmetries, which may help to construct practical photonic devices with robust optical functionalities.

\begin{figure}[tp]
\centerline{\includegraphics[width=8.8cm]{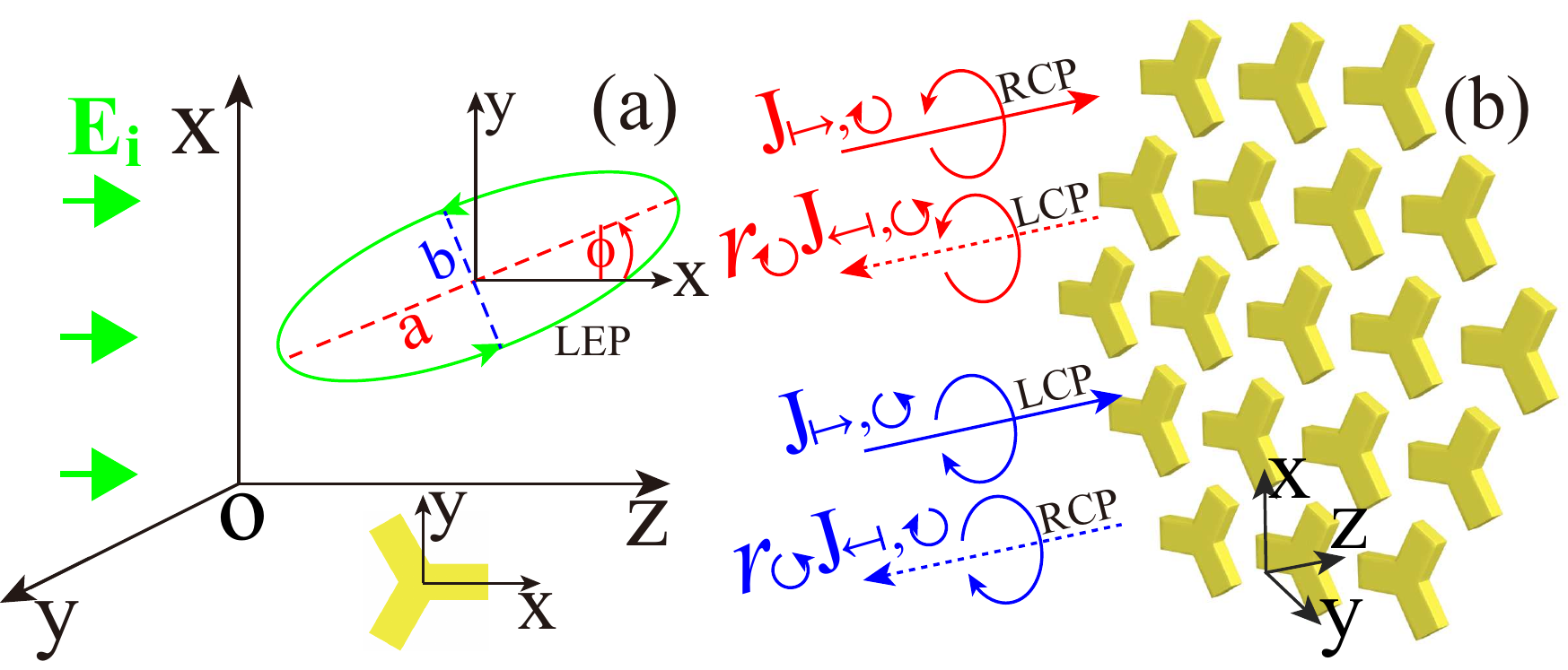}} \caption {\small (a) Plane wave (propagating along \textbf{+z} direction) scattering by rotationally-symmetric obstacles (symmetry axis parallel to \textbf{z}-axis). For generally elliptically-polarized incident waves, the polarization ellipse can be characterized by eccentricity $e=\pm b/a$ (the inset shows a left-handed ellipse with $e>0$) and orientation angle $\phi$. (b) Schematic  illustration for the backscattering or reflection processes for incident circularly-polarized waves. The backscattered or reflected waves are also circularly-polarized with opposite handedness.}\label{fig1}
\end{figure} 

\section{Theoretical analysis based on reciprocity and rotation symmetries}
\label{section2}

The scattering configuration we study is shown schematically in Fig.~\ref{fig1}(a): the incident plane wave is propagating along \textbf{+z} direction and the n-fold ($n\geq3$) rotation symmetry axis of the scattering body is also parallel to the \textbf{z} axis. Here we show only a $3$-fold rotationally-symmetric scatterer but note that the principles we reveal are applicable to all scenarios of $n\geq3$. For the analysis in this study: $\mymapstor$ and $\mymapstol$ denote waves propagating along \textbf{+z} and \textbf{-z} directions, respectively; $\circlearrowleft$ and $\circlearrowright$ denote waves of left-handed circularly-polarized (LCP) and right-handed circularly-polarized (RCP) waves, respectively (viewed counter to the direction of propagation). With the temporal phase component $\exp(-i\omega t)$ dropped, the normalized complex Jones vectors for the four possible combinations are~\cite{YARIV_2006__Photonics}:
\begin{equation}
\label{jones-vector}
\begin{aligned}
\mathbf{J}_{\mymapstor, \circlearrowleft} &=\frac{1}{\sqrt{2}}\left(\begin{array}{c}
1 \\
i
\end{array}\right) e^{i k z}, ~ \mathbf{J}_{\mymapstol, \circlearrowleft}=\frac{1}{\sqrt{2}}\left(\begin{array}{c}
1 \\
-i
\end{array}\right) e^{-i k z}; \\
\mathbf{J}_{\mymapstor, \circlearrowright} &=\frac{1}{\sqrt{2}}\left(\begin{array}{c}
1 \\
-i
\end{array}\right) e^{i k z},~ \mathbf{J}_{\mymapstol, \circlearrowright}=\frac{1}{\sqrt{2}}\left(\begin{array}{c}
1 \\
i
\end{array}\right) e^{-i k z}.
\end{aligned}\end{equation}

For circularly-polarized incident waves of $\mathbf{J}_{\mymapstor, \circlearrowleft}$ and $\mathbf{J}_{\mymapstor, \circlearrowright}$, the n-fold ($n\geq3$) rotation symmetry of the scattering body ensures that the backscattered waves are also circularly-polarized of the opposite handedness~\cite{FERNANDEZ-CORBATON_2013_Opt.ExpressOE_Forwarda,YANG_2020_ArXiv200613466Phys._Symmetry}. The backscattering processes can be expressed as [also shown schematically in Fig.~\ref{fig1}(b)]:
\begin{equation}
\label{r-coefficient-circular}
\begin{aligned}
\mathbf{J}_{\mymapstor, \circlearrowleft} \Rightarrow  r_\circlearrowleft\mathbf{J}_{\mymapstol, \circlearrowright};\\
\mathbf{J}_{\mymapstor, \circlearrowright} \Rightarrow r_\circlearrowright\mathbf{J}_{\mymapstol, \circlearrowleft},
\end{aligned}
\end{equation}
where $r_\circlearrowleft$ and $r_\circlearrowright$ denote backscattering coefficient for incident LCP and RCP waves, respectively. Equation~(\ref{r-coefficient-circular}) basically tells that upon backscattering both the propagation direction and the handedness of the wave are flipped.

Similar to the backscattering process for rotationally-symmetric scatterers, a time reversal operation (\textbf{\^{T}}) also reverses the wave propagation direction, but maintains the wave handedness:
\begin{equation}
\label{time-reversal}
\mathbf{\hat{T}}:\mathbf{J}_{\mymapstor, \circlearrowleft}
\Leftrightarrow \mathbf{J} _{\mymapstol, \circlearrowleft}, ~~~\mathbf{J}_{\mymapstor, \circlearrowright}
\Leftrightarrow  \mathbf{J}_{ \mymapstol, \circlearrowright},
\end{equation}
which in our notations of this work is effectively equivalent to the complex conjugate operation ($^*$), since according to Eq.~(\ref{jones-vector}):
\begin{equation}
\label{complex-conjugate}
(\mathbf{J}_{\mymapstor, \circlearrowleft})^*
= \mathbf{J} _{\mymapstol, \circlearrowleft}, ~~~(\mathbf{J}_{\mymapstor, \circlearrowright})^*
=  \mathbf{J}_{ \mymapstol, \circlearrowright}.
\end{equation}
The principle of electromagnetic reciprocity~\cite{POTTON_Rep.Prog.Phys._reciprocity_2004} requires that the incident and backscattered circularly-polarized field components have to satisfy the following equation:
\begin{equation}
\label{reciprocity-1}
(\mathbf{\hat{T}}\mathbf{J}_{\mymapstor, \circlearrowright})^\dag r_\circlearrowleft\mathbf{J}_{\mymapstol, \circlearrowright}=(\mathbf{\hat{T}}\mathbf{J}_{\mymapstor, \circlearrowleft})^\dag r_\circlearrowright\mathbf{J}_{\mymapstol, \circlearrowleft},
\end{equation}
which according to Eqs.~(\ref{time-reversal}) and (\ref{complex-conjugate}) is effectively:
\begin{equation}
\label{reciprocity}
(\mathbf{J}_{\mymapstor, \circlearrowright}^*)^\dag r_\circlearrowleft\mathbf{J}_{\mymapstol, \circlearrowright}=(\mathbf{J}_{\mymapstor, \circlearrowleft}^*)^\dag r_\circlearrowright\mathbf{J}_{\mymapstol, \circlearrowleft},
\end{equation}
where $^\dag$ corresponds to the complex conjugate-transpose operation: $\binom{\alpha}{\beta}^\dag=(\alpha^*,\beta^*)$ ($\alpha$ and $\beta$ are arbitrary complex numbers). According to Eq.~(\ref{complex-conjugate}), Eq.~(\ref{reciprocity}) can be reduced to ($r_\circlearrowleft$ and $r_\circlearrowright$ are complex scalars):
\begin{equation}
\label{reciprocity-reduced}
r_\circlearrowleft(\mathbf{J}_{\mymapstol, \circlearrowright})^\dag\mathbf{J}_{\mymapstol, \circlearrowright}=r_\circlearrowright(\mathbf{J}_{\mymapstol, \circlearrowleft})^\dag \mathbf{J}_{\mymapstol, \circlearrowleft}.
\end{equation}
It is easy to confirm from Eq.~(\ref{jones-vector}) that $(\mathbf{J}_{\mymapstol, \circlearrowright})^\dag\mathbf{J}_{\mymapstol, \circlearrowright}=(\mathbf{J}_{\mymapstol, \circlearrowleft})^\dag \mathbf{J}_{\mymapstol, \circlearrowleft}=1$ (this is effectively the definition for squared length of the normalized complex Jones vectors), as a result we obtain circular polarization-independent backscattering coefficient and efficiency:
\begin{equation}
\label{coefficient-efficiency}
r_\circlearrowleft=r_\circlearrowright=r_c, ~~~ R_\circlearrowleft=R_\circlearrowright=|r_c|^2.
\end{equation}

After studying circular polarizations, now we turn to the scenario of arbitrarily-polarized incident waves, which can always be expanded into the circular basis as follows (in terms of the electric field):
\begin{equation}
\label{arbitrary incident}
\mathbf{E}_i=c_\circlearrowleft\mathbf{J}_{\mymapstor, \circlearrowleft}+c_\circlearrowright\mathbf{J}_{\mymapstor, \circlearrowright}.
\end{equation}
Here the incident field intensity is normalized:  $|c_\circlearrowleft|^2+|c_\circlearrowright|^2=1$. We further define the circular polarization component ratio  $c=c_\circlearrowright/c_\circlearrowleft$, where obviously $c=0$ and $c=\infty$ corresponds to LCP and RCP waves, respectively. For other general elliptical polarizations [refer to the inset of Fig.~\ref{fig1}(a), where a left-handed elliptically-polarized (LEP) wave is shown]: $|c|$ decides the eccentricity $e=\pm b/a$ (the positive and negative sign corresponds to left and right handedness, respectively) and thus also handedness of the polarization ellipse; its phase Arg($c$) decides the orientation angle $\phi$ of the polarization ellipse (the angle between the \textbf{x}-axis and the semi-major axis of the ellipse)~\cite{YANG_2020_ArXiv200613466Phys._Symmetry,YANG_ArXiv200701535Phys._Scatteringa}. For example, $|c|<1$, $|c|>1$ and $|c|=1$ correspond respectively to left-handed, right-handed and linear polarizations. According to Eq.~(\ref{r-coefficient-circular}), the backscattered field is:
\begin{equation}
\label{arbitrary-reflection}
\mathbf{E}_r=r_c(c_\circlearrowleft \mathbf{J}_{\mymapstol, \circlearrowright}+c_\circlearrowright \mathbf{J}_{\mymapstol, \circlearrowleft}).
\end{equation}
Since the two reflected components $\mathbf{J}_{\mymapstol, \circlearrowright}$ and $\mathbf{J}_{\mymapstol, \circlearrowleft}$ are orthogonal, the backscattering coefficient and efficiency for arbitrary polarizations can be expressed as:
\begin{equation}\begin{array}{c}
r_{a}=\frac{\mathbf{E}_{r}}{\mathbf{E}_{i}}=r_{c};\\
R_{a}=\frac{\left|\mathbf{E}_{r}\right|^{2}}{\left|\mathbf{E}_{i}\right|^{2}}=\left|r_{c}\right|^{2}\left(\left|c_{\circlearrowleft}\right|^{2}+\left|c_{\circlearrowright}\right|^{2}\right)=\left|r_{c}\right|^{2},
\end{array}\end{equation}
which obviously have nothing to do with the polarization state of the incident plane wave.

We note here that our discussions above are fully based on the fundamental principle of reciprocity and geometric rotation symmetry, which can be directly extended from finite scattering bodies to infinite periodic or even quasi-periodic structures. Then $r_{a}$ and $R_{a}$ represent respectively reflection coefficient and efficiency, which in a similar fashion are also arbitrary polarization-independent. Despite this similarity, there are also obvious differences between finite scattering bodies and infinite periodic structures. Besides the common absorption channel:  for the former there are an infinite number of out-coupling channels that correspond to scatterings along all directions on the momentum sphere; while for the latter, the number of the out-coupling channels are finite, including reflection, transmission and other possible diffractions limited by the periodicity. Without absorption (no Ohmic losses) or diffractions (periodicity is small enough to make all diffractions evanescent), rotation symmetries would directly lead to the invariance of both reflections and transmissions, as is required by the law of energy conservation.

\section{Invariant backscattering of finite scatterers}
\label{section3}

\begin{figure}[tp]
\centerline{\includegraphics[width=8cm]{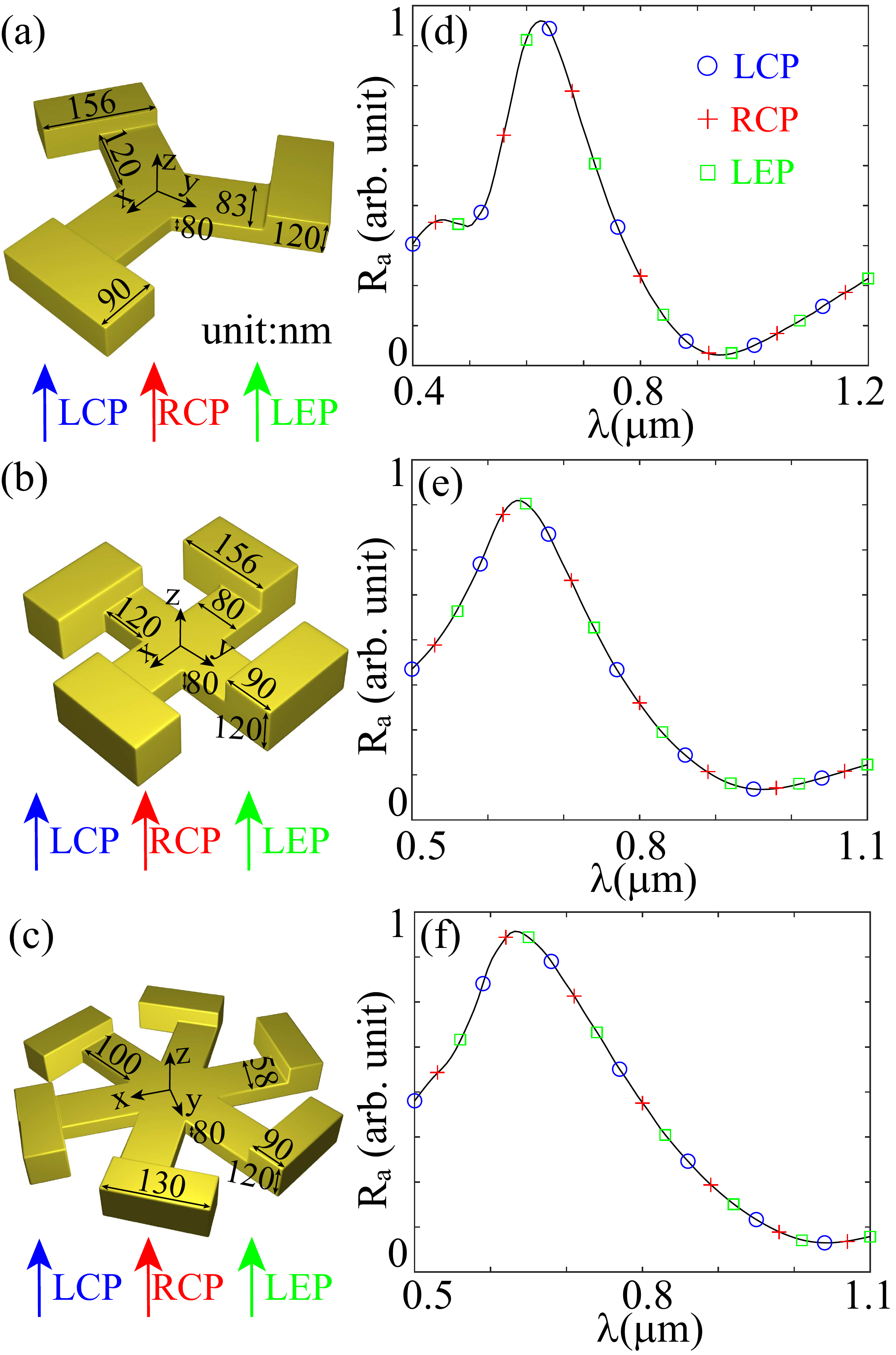}} \caption{\small  (a)-(c) Finite scattering bodies made of gold that exhibit $3$-fold, $4$-fold and $6$-fold rotation symmetries, respectively. The geometric parameters are specified in the figure. (d)-(f) The corresponding backscattering efficiency spectra for the structures in (a)-(c). For each structure, three sets of spectra are shown for three incident polarizations:  LCP, RCP, and LEP ($e=0.5$ and $\phi=45^\circ$). }
\label{fig2}
\end{figure}

To verify our theoretical analysis above, firstly we investigate finite scattering bodies and the numerically-calculated results are shown in Fig.~\ref{fig2} (numerical results in this work are obtained through commercial software COMSOL Multiphysics).  Here we have studied three gold  particles (optical parameters taken from Ref.~\cite{Johnson1972_PRB}; geometric parameters specified in the figure) that exhibit $3$-fold, $4$-fold and $6$-fold rotation symmetries respectively [see Figs.~\ref{fig2}(a)-(c); no other geometric symmetries are shown].  The corresponding backscattering efficiency are shown respectively in Figs.~\ref{fig2}(d)-(f), and for each scenario we show three sets of spectra with different incident polarizations: two circular polarizations of opposite handedness (LCP and RCP) and one LEP with $e=0.5$ and $\phi=45^\circ$. We emphasize that in this study we have only randomly chosen three
representative incident polarizations and the results are the same for other polarizations as well, as is the case throughout our work. Moreover, the backscattering invariance shown here can be also obtained for any $n$-fold ($n\geq3$) rotationally-symmetric scatterers, though here we have showcased only three scenarios. Those results have confirmed our conclusion that sole rotation symmetries can ensure arbitrary polarization-independent backscattering by reciprocal scattering bodies.

\section{Invariant optical responses of periodic structures}

\begin{figure}[tp]
\centerline{\includegraphics[width=8.9cm]{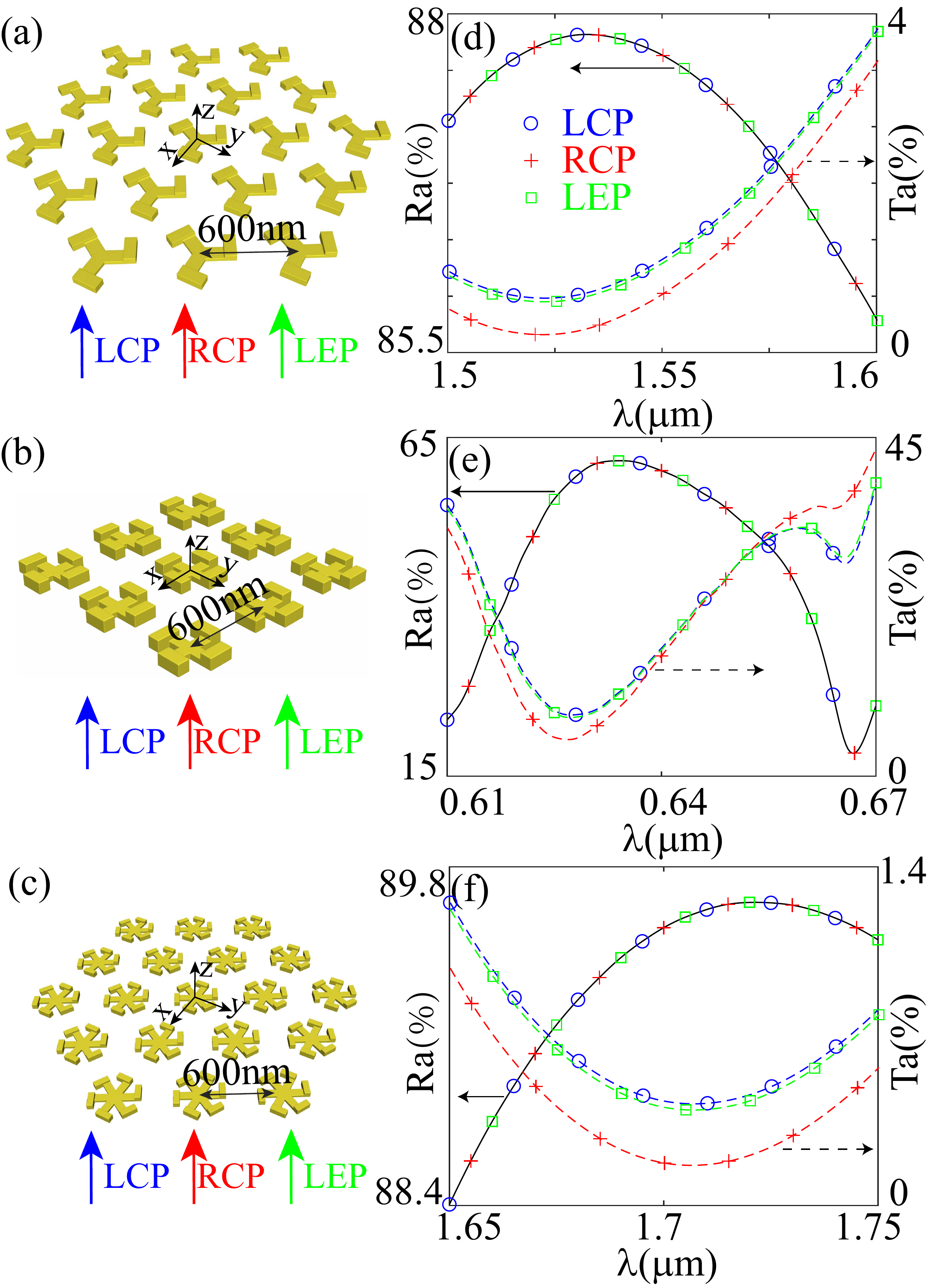}} \caption{\small  (a)-(c) Periodic structures [the constituent unit-cell particles are the same as those in  Figs.~\ref{fig2}(a)-(c), respectively] that exhibit $3$-fold, $4$-fold and $6$-fold rotation symmetries, respectively. There are two triangular lattices in (a) and (c), and one square lattice in (b). The periodicity parameters are specified in the figure.  (d)-(f) The corresponding transmission and reflection efficiency spectra for the structures in (a)-(c).}\label{fig3}
\end{figure}

\subsection{Invariant reflections induced by rotation symmetries}
As a next step, we turn to extended periodic structures that show $3$-fold, $4$-fold and $6$-fold rotation symmetries respectively [see Figs.~\ref{fig3}(a)-(c), where there are two triangular lattices in Figs.~\ref{fig3} (a) and (c), and one square lattice in Fig.~\ref{fig3}(b)]: the consisting unit-cell particles are the same as those in Figs.~\ref{fig2}(a)-(c), respectively; and the periodicity information is specified in the figure. The corresponding reflection spectra are shown respectively in Figs.~\ref{fig3}(d)-(f), where their independence on the polarization is clearly observed. Besides the reflection spectra, we have also included the results for transmission efficiency spectra of arbitrary polarizations ($T_a$), for which in contrast there is no such polarization independence.  In the spectral regime we study there are no other diffractions and consequently this variance of transmission with respect to polarizations must be induced by the Ohmic losses of gold.  The law of energy conservation guarantees the coexistence of  transmission and reflection invariance when there are neither Ohmic losses nor extra diffractions.

\begin{figure}[tp]
\centerline{\includegraphics[width=8.5cm]{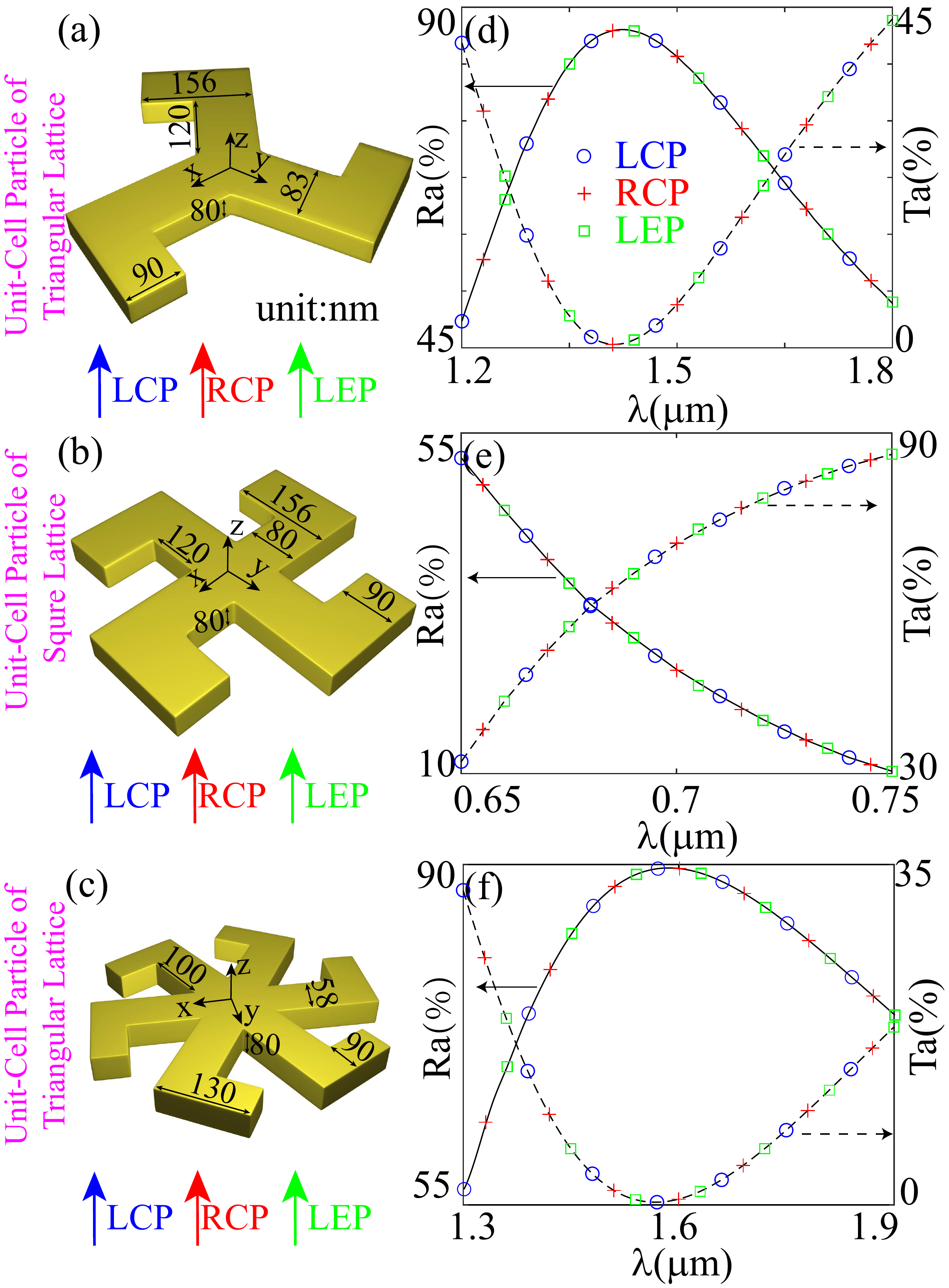}} \caption{\small \small  (a)-(c) Periodic structures [only the unit-cell particles are shown; the periodicity parameters and rotation symmetries are the same as those in  Figs.~\ref{fig3}(a)-(c), respectively] that also exhibit extra reflection symmetry with the reflection mirror perpendicular to the incident direction. The geometric parameters of the unit-cell particles are specified in the figure. (d)-(f) The corresponding transmission and reflection efficiency spectra for the structures in (a)-(c).}
\label{fig4}
\end{figure}

We emphasize here that the reflection invariance for some special scenarios (such as $3$-fold and $4$-fold rotationally-symmetric lattices with circularly-polarized incident waves) has already been demonstrated in previous studies~\cite{KASCHKE_Opt.Express_metamaterial,KASCHKE_Opt.Express_Tapered,KONDRATOV_Phys.Rev.B_Extreme,GORKUNOV_Phys.Rev.Lett._Metasurfaces}, where relatively more algebraic arguments centered on the scattering matrix have been put forward to clarify the mechanisms. In comparison, our arguments in this study are more geometric and intuitive, and mostly importantly they are all-inclusive, universally covering all $n$-fold ($n\geq3$) rotational symmetries and all polarizations. Besides, our conclusions are valid for both finite scattering bodies and extended periodic (or quasi-periodic) structures,  involving no complicated algebraic manipulations of the scattering matrices.

\subsection{Invariant reflections and transmissions induced by joint rotation-reflection symmetries}
\begin{figure}[tp]
\centerline{\includegraphics[width=8.5cm]{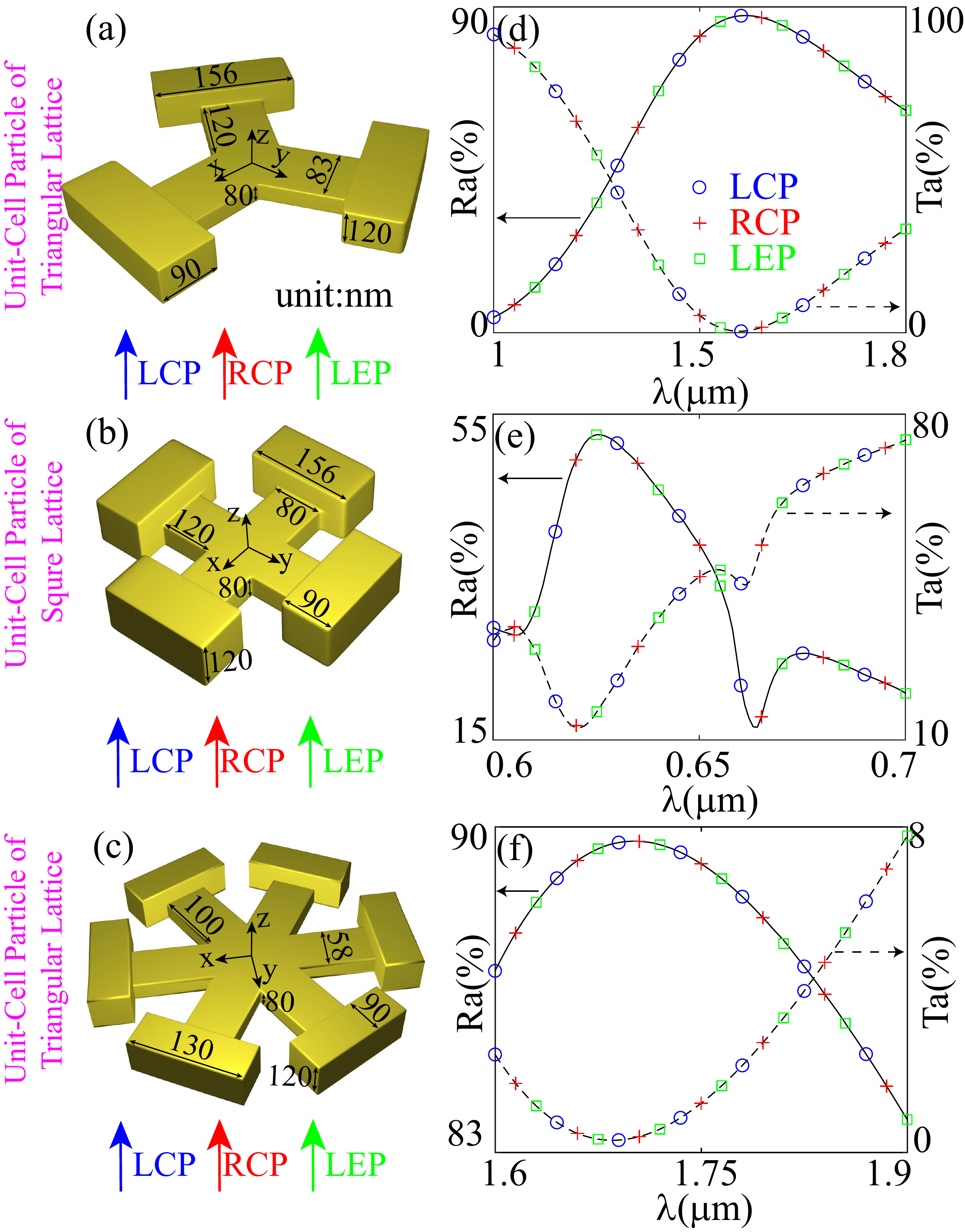}} \caption{\small (a)-(c) Periodic structures [only the unit-cell particles are shown; the periodicity parameters and rotation symmetries are the same as those in  Figs.~\ref{fig3}(a)-(c), respectively] that also exhibit extra reflection symmetry with the reflection mirror parallel to the incident direction. The geometric parameters of the unit-cell particles are specified in the figure. (d)-(f) The corresponding transmission and reflection efficiency spectra for the structures in (a)-(c).}
\label{fig5}
\end{figure}

Now we proceed to discuss how to obtain both invariance of reflection and transmission at the presence of losses.  Our previous studies have shown that $n$-fold ($n\geq3$) rotation symmetries together with extinction invariance for unit-cell scattering bodies can lead to transmission invariance~\cite{YANG_2020_ArXiv200613466Phys._Symmetry,CHEN_2020_Phys.Rev.Research_Scatteringa}.  The extinction invariance for arbitrary polarizations can be achieved by introducing extra reflection symmetries, with the reflection plane either  perpendicular or parallel to the incident direction (perpendicular or parallel reflection symmetry)~\cite{YANG_2020_ArXiv200613466Phys._Symmetry,CHEN_2020_Phys.Rev.Research_Scatteringa}.  The results of the joint rotation-perpendicular (parallel) reflection symmetry scenario are summarized in Fig.~\ref{fig4} (Fig.~\ref{fig5}).  The periodic structures (two triangular lattices and one square lattice for each scenario; only unit-cell particles are shown) in Figs.~\ref{fig4}(a)-(c) and Figs.~\ref{fig5}(a)-(c) are the same as those in Figs.~\ref{fig3}(a)-(c), except now that the unit-cell particles are modified to accommodate the corresponding perpendicular or parallel reflection symmetry.  The corresponding reflection and transmission spectra in Figs.~\ref{fig4}(d)-(f) and Figs.~\ref{fig5}(d)-(f) have verified the invariance of both reflections and transmissions for arbitrary polarizations. When there are no extra diffractions (as is the case for our studies in Fig.~\ref{fig4}), it is known from the law of energy conservation that the invariance of reflections and transmissions means also absorption invariance. For the parallel reflection symmetry scenario shown in Fig.~\ref{fig5},  the parity conservation  and rotation symmetry directly guarantee the invariance of absorption, irrespective of whether there are higher-order diffractions or not~\cite{YANG_2020_ArXiv200613466Phys._Symmetry,CHEN_2020_Phys.Rev.Research_Scatteringa}.

We further note that the principles we have revealed concerning the reflection and transmission invariance are secured by the fundamental laws of reciprocity and parity conservation, irrespective of whether or not there are extra diffractions~\cite{YANG_2020_ArXiv200613466Phys._Symmetry,CHEN_2020_Phys.Rev.Research_Scatteringa}. Of course, at the presence of higher order diffractions, we have to specify that what we mean by reflection and transmission are actually the corresponding zero-order diffractions. At the same time, the invariance of reflection and transmission does not necessarily result in absorption invariance, since there are other out-coupling diffraction channels.

\section{Absence of backscattering or reflection invariance for self-dual optical structures}
As a final step, we turn to self-dual photonic structures that are invariant under the electromagnetic duality transformation~\cite{JACKSON_1998__Classical,FERNANDEZ-CORBATON_2013_Phys.Rev.Lett._Electromagnetica,YANG_2020_ACSPhotonics_Electromagnetic,YANG_ArXiv200701535Phys._Scatteringa,FERNANDEZ-CORBATON_2013_Opt.ExpressOE_Forwarda}. This is because self-dual and rotationally-symmetric structures are similar in the sense that for incident circularly-polarized waves, the backscattered or reflected waves are also circularly-polarized (actually for self-dual ones, along all scattering or out-coupling directions the waves are either circularly-polarized or zero)~\cite{FERNANDEZ-CORBATON_2013_Phys.Rev.Lett._Electromagnetica,YANG_ArXiv200701535Phys._Scatteringa,FERNANDEZ-CORBATON_2013_Opt.ExpressOE_Forwarda}. The contrasting difference is that the handedness of the waves upon backscattering or reflection is preserved for self-dual structures, while flipped for rotationally-symmetric one. As a result, when the formula and analysis in Section~\ref{section2} are mapped to self-dual structures, Eq.~(\ref{reciprocity-reduced}) would be converted to:
\begin{equation}
\label{reciprocity-reduced-dual}
r_\circlearrowleft(\mathbf{J}_{\mymapstol, \circlearrowright})^\dag\mathbf{J}_{\mymapstol, \circlearrowleft}=r_\circlearrowright(\mathbf{J}_{\mymapstol, \circlearrowleft})^\dag \mathbf{J}_{\mymapstol, \circlearrowright}.
\end{equation}
According to Eq.~(\ref{jones-vector}), $(\mathbf{J}_{\mymapstol, \circlearrowright})^\dag\mathbf{J}_{\mymapstol, \circlearrowleft}=(\mathbf{J}_{\mymapstol, \circlearrowleft})^\dag \mathbf{J}_{\mymapstol, \circlearrowright}=0$ (this is effectively the definition of orthogonality between complex Jones vectors that characterize waves of opposite handedness while the same propagation direction), and consequently Eq.~(\ref{reciprocity-reduced-dual}) is generically satisfied, telling nothing about the backscattering coefficients. This to say, in sharp contrast to the rotationally-symmetric scenario, self-duality does not necessarily lead to backscattering even for circular polarizations. \\

\begin{figure}[tp]
\centerline{\includegraphics[width=8cm]{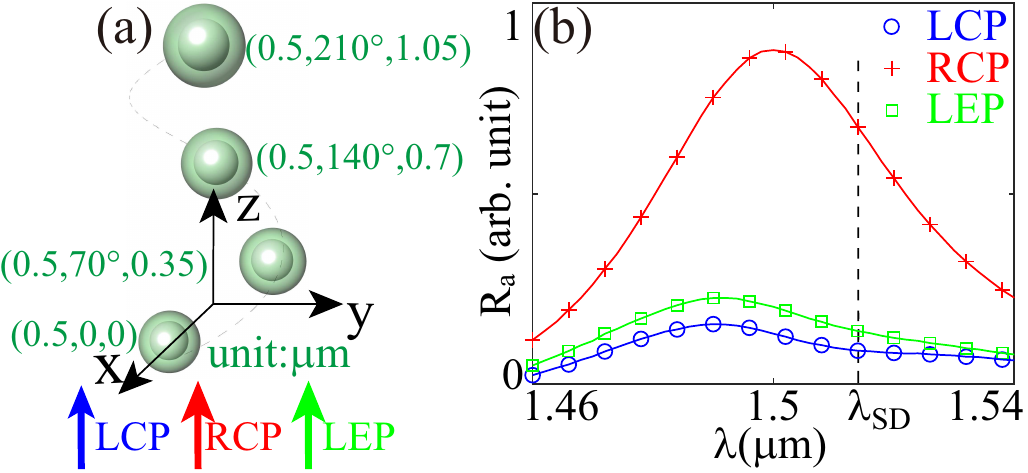}} \caption{\small (a) The  scattering configuration consisting of four self-dual (at ${\lambda_\mathbf{SD}}$) Ag core-dielectric ($n=3.4$) shell spherical particles.  The cylindrical coordinates of the particles are specified in the figure and this configuration exhibits neither rotation nor reflection symmetry.  (b) The corresponding backscattering efficiency spectra for three incident polarizations, with the self-dual wavelength marked by a dashed vertical line.}
\label{fig6}
\end{figure}

To verify the above analysis, we restudy the self-dual scattering cluster studied in Ref.~\cite{YANG_ArXiv200701535Phys._Scatteringa}, but now from a different perspective of backscattering efficiency rather than total scattering cross sections. The cluster consist of Ag core-dielectric shell spherical particle (inner radius $65$~nm and outer radius $218$~nm; the refractive index of Ag is adopted from Ref.~\cite{Johnson1972_PRB} and that of the dielectric shell is $n=3.4$), which is self-dual at the wavelength ${\lambda_\mathbf{SD}} = 1512$~nm (a pair of electric and magnetic dipoles of equal strength are supported at ${\lambda_\mathbf{SD}}$).  The scattering configuration is shown schematically in Fig.~\ref{fig6}(a), where the cylindrical coordinates ($r=\sqrt{x^2+y^2},\phi,z$) of particles are also specified. The backscattering efficiency spectra are summarized in Fig.~\ref{fig6}(b), which clearly demonstrate that self-duality at ${\lambda_\mathbf{SD}}$ does not lead to invariance of the backward scattering. \\

\section{Conclusion}

To conclude, we reveal that for reciprocal finite scattering bodies or extended periodic structures, rotation symmetries (no less than three-fold) are sufficient to guarantee invariant backscattering or reflection for incident plane waves of arbitrary polarizations. For periodic structures, it is further shown that the incorporation of extra reflection symmetries (with the reflection plane either perpendicular or parallel to the incident direction) would also secure the polarization independence of transmissions. It means invariant reflections, transmissions, and absorptions when there are no extra higher-order diffraction orders. For numerical demonstrations in Figs.~\ref{fig3}-\ref{fig6}, we employ only periodic structures that exhibit  $3$-fold, $4$-fold and $6$-fold rotation symmetries, but emphasize that the conclusions are also valid for quasi-periodic structures that harbor other rotation symmetries.  The principles we have revealed are generically protected by fundamental laws of reciprocity, parity conservation and rotation symmetries, which may play significant roles in both fundamental explorations involving polarization states of light (such as investigations in chiroptics and the polarization evolving Pancharatnam-Berry phase~\cite{berry_adiabatic_1987}) and practical applications that need stable optical functionalities that are robust against polarization fluctuations.

\section*{acknowledgement}
We acknowledge the financial support from National Natural Science
Foundation of China (Grant No. 11874026 and 11874426), and the Outstanding Young Researcher Scheme of National University of Defense Technology. W. L. thanks
Yuri S. Kivshar and Maxim V. Gorkunov for useful correspondences.



\end{document}